\documentclass{article}

\usepackage{arxiv}

\usepackage[utf8]{inputenc} % allow utf-8 input
\usepackage[T1]{fontenc}    % use 8-bit T1 fonts
\usepackage{hyperref}       % hyperlinks
\usepackage{url}            % simple URL typesetting
\usepackage{booktabs}       % professional-quality tables
\usepackage{amsfonts}       % blackboard math symbols
\usepackage{nicefrac}       % compact symbols for 1/2, etc.
\usepackage{microtype}      % microtypography

\usepackage{tabularx}
\usepackage{multirow}
\usepackage{amsthm}
\usepackage{amsmath, amsfonts}
\usepackage{tikz}
\usepackage{pgfplots}
\pgfplotsset{compat=1.18}
\usepackage{enumitem}
\usepackage{soul}
\usepackage{mwe}
\usepackage{booktabs}

\usepackage{cleveref}       % smart cross-referencing
\usepackage{lipsum}         % Can be removed after putting your text content
\usepackage{graphicx}
\usepackage{natbib}
\usepackage{doi}

\title{PathGPT: Reframing Path Recommendation as a Natural Language Generation
Task with Retrieval-Augmented Language Models}

\date{August 20, 2025}
\newif\ifuniqueAffiliation
\uniqueAffiliationtrue

\ifuniqueAffiliation % Standard variant of author block
\author{ \href{https://orcid.org/0009-0003-8107-9274}               {\includegraphics[scale=0.06]{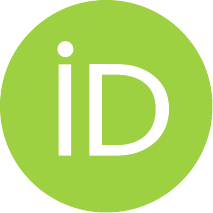}\hspace{1mm}Steeve C.~Marcelyn}\\
	Department of Computer Science\\
	Shanghai Jiao Tong University\\
	Shanghai, China\\
	\texttt{stephen\_5@sjtu.edu.cn} \\
	\And
	Yucen~Gao \\
	Department of Computer Science\\
	Shanghai Jiao Tong University\\
	Shanghai, China\\
	\texttt{guo\_ke@sjtu.edu.cn} \\
        \And
        Yuzhe~Zhang \\
	Department of Computer Science\\
	Shanghai Jiao Tong University\\
	Shanghai, China\\
	\texttt{zhangyuzhe@sjtu.edu.cn} \\
        \And
        Xiaofeng~Gao \\
	Department of Computer Science\\
	Shanghai Jiao Tong University\\
	Shanghai, China\\
	\texttt{gao-xf@cs.sjtu.edu.cn} \\
}
\else
\usepackage{authblk}

\newbox{\orcid}\sbox{\orcid}{\includegraphics[scale=0.06]{orcid.pdf}} 
\author[1]{%
	\href{https://orcid.org/0000-0000-0000-0000}{\usebox{\orcid}\hspace{1mm}David S.~Hippocampus\thanks{\texttt{hippo@cs.cranberry-lemon.edu}}}%
}
\author[1,2]{%
	\href{https://orcid.org/0000-0000-0000-0000}{\usebox{\orcid}\hspace{1mm}Elias D.~Striatum\thanks{\texttt{stariate@ee.mount-sheikh.edu}}}%
}
\affil[1]{Department of Computer Science, Cranberry-Lemon University, Pittsburgh, PA 15213}
\affil[2]{Department of Electrical Engineering, Mount-Sheikh University, Santa Narimana, Levand}
\fi

\renewcommand{\headeright}{}
\renewcommand{\undertitle}{}
\renewcommand{\shorttitle}{PathGPT: Reframing Path Recommendation as a Natural Language Generation Task with Retrieval-Augmented LLMs}

\hypersetup{
pdftitle={PathGPT: Reframing Path Recommendation as a Natural Language Generation Task with Retrieval-Augmented Language Models},
pdfsubject={cs.IR},
pdfauthor={Steeve C.~Marcelyn, Yucen~Gao, Yuzhe~Zhang, Xiaofeng~Gao},
pdfkeywords={Path Recommendation, Deep Learning, Large Language Model, Retrieval Augmented Generation, Natural Language Processing},
}

\newtheorem{definition}{Definition}

\begin{document}
\maketitle

\begin{abstract}
Path recommendation (PR) aims to generate travel paths that are customized to a user's specific preferences and constraints. Conventional approaches often employ explicit optimization objectives or specialized machine learning architectures; however, these methods typically exhibit limited flexibility and generalizability, necessitating costly retraining to accommodate new scenarios. This paper introduces an alternative paradigm that conceptualizes PR as a natural language generation task. We present PathGPT, a retrieval-augmented large language model (LLM) system that leverages historical trajectory data and natural language user constraints to generate plausible paths. The proposed methodology first converts raw trajectory data into a human-interpretable textual format, which is then stored in a database. Subsequently, a hybrid retrieval system extracts path-specific context from this database to inform a pretrained LLM. The primary contribution of this work is a novel framework that demonstrates how integrating established information retrieval and generative model components can enable adaptive, zero-shot path generation across diverse scenarios. Extensive experiments on large-scale trajectory datasets indicate that PathGPT's performance is competitive with specialized, learning-based methods, underscoring its potential as a flexible and generalizable path generation system that avoids the need for retraining inherent in previous data-driven models. Our code is available at \url{https://github.com/Kuramenai/PathGPT/}.
\end{abstract}

% keywords can be removed

\section{Introduction}
\emph{Path recommendation} is one of the core services of many GPS-aided navigation systems~\cite{neuromlr}. And conventionally, graph-based routing algorithms, such as Dijkstra's and A*, are frequently used to find and recommend travel paths, often the fastest or the shortest, to users navigating through increasingly complex road networks~\cite{increase_net}. However, empirical analysis of large-scale trajectory data reveals a significant discrepancy: human drivers may sometimes choose paths that are neither the fastest nor the shortest~\cite{personal_route}~\cite{driver_behavior}. This divergence highlights a fundamental limitation of conventional search-based algorithms, as real-world navigation involves multiple, often implicit factors that are difficult to encode, such as prioritizing a scenic view over travel time or avoiding highways during a leisurely drive~\cite{leisure_drive}. Recent advances in computer hardware, combined with the availability of large-scale trajectory datasets~\cite{gps_data_1}, have spurred the development of data-driven path recommendation algorithms~\cite{prr_1}. Leveraging the ability of machine learning to identify latent patterns in data, these models can generate paths optimized for a variety of factors beyond simple distance, such as traffic conditions or proximity to points of interest, thereby offering more sophisticated solutions to the PR problem.

Many of these approaches can be viewed as learning a Markovian model, where the goal is to predict the next road segment based on the current one, given a transition probability matrix learned from historical data and user preferences. Provided a query $q$, the recommended path $\mathcal{P}$ can be obtained from the following equation:
\begin{equation}
P(\mathcal{P} | q) \approx \prod_{i=1}^{|\mathcal{P} |} P(e_i | e_{i-1}, q)
\label{eq:markov}
\end{equation}
However, despite their advancements, these data-driven models suffer from a significant limitation: a lack of flexibility. They are typically trained to generate paths based on a predetermined set of constraints, and the learned transition probability matrix cannot be easily dynamically adapted to new requirements post-training. For example, a model trained to optimize for traffic conditions cannot, without modification, generate a path that must pass through a specific sequence of user-defined points of interest. A naive solution to this inflexibility would be to train a distinct model for every conceivable user requirement. However, such an approach is not only computationally inefficient but also practically infeasible. The set of all potential user constraints is vast and often unknown a priori, making it impossible to develop a comprehensive suite of specialized models that could constitute a truly unified solution. Inspired by the remarkable success of Large Language Models (LLMs) across a diverse range of tasks, including question answering and code generation~\cite{brown2020languagemodelsfewshotlearners}, we postulate that the extensive world knowledge and sophisticated natural language understanding capabilities inherent in these models can be harnessed for path generation. We also note that the autoregressive nature of LLMs,  which is encapsulated by the next token prediction (Equation~\ref{eq:llm}), is also expressed in a similar form to Equation~\ref{eq:markov}, which may imply that although representing different architectures, they might share some similarities in their modeling.
\begin{equation}
P(x_1, x_2, \dots, x_T) = \prod_{t=1}^{T} P(x_t \mid x_1, x_2, \dots, x_{t-1})
\label{eq:llm}
\end{equation}

More importantly, the LLMs' fundamental architecture would enable conditioning the entire generation process on arbitrary user preferences expressed in natural language, a capability far beyond that of standard Markovian approaches. This inherent flexibility is what could make LLMs uniquely suited to solve the personalized path recommendation problem. Such an approach would enable the recommendation of paths finely tuned to specific user requirements. This paradigm could directly address the limitations of prior methods by avoiding the need for multiple, specialized models and adeptly handling the inherent variety and ambiguity of user queries.
Motivated by these findings, we developed PathGPT, a framework that enhances the path recommendation capabilities of LLMs by reformulating the Path Recommendation (PR) problem as a natural language processing task. The objective is for the LLM to generate a path that adheres to an origin-destination (OD) pair and a set of constraints, all of which are expressed in natural language. To ground the LLM's generation in factual trajectory data, our methodology is a form of Retrieval-Augmented Generation (RAG) and consists of two primary stages. First, we preprocess a database of historical paths, which are initially represented as sequences of road network edge IDs. This process transforms them into natural language by: 1) using reverse geocoding to obtain human-readable addresses for the OD nodes, and 2) translating the edge sequence into the corresponding traversed road names. This information is then synthesized into a coherent textual description for each path (see Figure \ref{fig:context_generation}) and indexed in a knowledge base. Second, during the generation phase, given a user query, a hybrid retriever then extracts relevant historical paths from the knowledge base based on the query's OD pair. These retrieved textual paths, combined with the original query, serve as rich, in-context information that guides the LLM in generating a final tailored path. 

%\st{The decision to use a textual representation is deliberate. While prior studies suggest LLMs have some capacity to interpret raw GPS coordinates, our analysis showed that their probabilistic nature and tokenization schemes lead to inconsistent outputs for such data. Given the inherent strength of LLMs in natural language tasks, our text-based approach ensures greater robustness and reliability in the generated paths.}~\cite{geollm}%

To the best of our knowledge, PathGPT is the first framework to apply LLMs to the PR problem by combining established information retrieval and generative model components. Our experiments demonstrate its unique capability to generate paths that accommodate novel user constraints at inference time, a task that poses a significant challenge for conventional data-driven models. Furthermore, PathGPT addresses the common 'black box' problem associated with many machine learning models. By leveraging the LLM's generative nature, the framework can produce natural language justifications for its recommended paths, thereby enhancing user trust and confidence in the system.

Our key contributions are as follows: 

\begin{itemize}
\item We propose a novel formulation of the path recommendation (PR) task as a natural language generation problem, enabling flexible user-defined constraints without retraining or goal-specific tuning.
\item We present PathGPT, a modular and reproducible system that integrates retrieval-augmented generation (RAG) with historical trajectory data and geospatial context, and demonstrates its applicability to path recommendation.
\item We introduce a textual transformation pipeline that converts raw trajectory data into semantically rich natural language prompts, facilitating reasoning by large language models.
\item We evaluate PathGPT on multiple large-scale datasets and demonstrate that even without fine-tuning, it can generate competitive paths, achieving performance on par with task-specific models.
\item This work opens up an alternative line of investigation for general-purpose language agents in the domain of geospatial reasoning and intelligent routing while showcasing their feasibility for future deployment.
\end{itemize}

\section{Related Work}
\subsection{Path Recommendation Systems}
Research in path recommendation has evolved from classical graph-based algorithms to specialized data-driven models.
\begin{itemize}
    \item Classical Graph-Based Approaches: Foundational approaches to path recommendation are dominated by algorithms like Dijkstra's and A*~\cite{ref_Dijkstra, ref_A}. These methods are effective at finding an optimal path based on a single, static cost metric such as distance or time. However, their primary limitations are their computational complexity on large networks and their inability to incorporate dynamic, multifaceted user preferences. Numerous variants have been proposed to address specific shortcomings, such as optimizing for fuel consumption (e.g., IA*FC)~\cite{ref_IAFC} or adapting to specific environments (e.g., A*-DWA)~\cite{ref_A-DWA}, but they often remain constrained to a single, hard-coded objective.
    \item Data-Driven and Learning-Based Models: To capture the nuances of human behavior, the field shifted towards data-driven models that learn from historical trajectory data. Models like NeuroMLR~\cite{neuromlr} can learn to predict paths based on factors latent in the data, achieving strong performance on specific tasks. Despite their sophistication, these models share a fundamental limitation with their classical predecessors: a lack of adaptability. They are typically trained for a fixed set of predefined objectives (e.g., fastest path, least traffic) and cannot generalize to novel, user-specified constraints at inference time without costly retraining. 
    %\st{Our work addresses this critical gap in flexibility}.%
\end{itemize}

\subsection{Large Language Models for Reasoning and Planning}
In parallel with these specialized developments, the emergence of Large Language Models (LLMs) has introduced a new paradigm for general-purpose reasoning~\cite{ref_LLM}. Models such as those from OpenAI and DeepSeek have demonstrated remarkable capabilities across diverse domains~\cite{ref_deepseek}. This proficiency stems from two core advancements: 1) a deep understanding of natural language, enabling them to identify entities and causal relationships from unstructured text~\cite{ref_LLM_benefit1}, and 2) the underlying Transformer architecture, which facilitates combinatorial generalization and symbolic reasoning~\cite{ref_LLM_benefit2}.
This inherent capacity for reasoning about complex, natural language instructions makes LLMs a compelling candidate for overcoming the rigidity of traditional path recommendation systems. They offer the potential to interpret a wide array of user preferences and constraints directly, moving beyond fixed optimization targets.

\subsection{Retrieval-Augmented Generation for Grounded Reasoning }
Despite their impressive capabilities, off-the-shelf LLMs are not inherently grounded in real-time or domain-specific factual knowledge. Their internal knowledge is static up to their last training date, which can lead to outdated or incorrect information~\cite{ref_RAG1}, and they are prone to "hallucination," where they generate plausible but factually incorrect details~\cite{ref_RAG2}.
The Retrieval-Augmented Generation (RAG) framework was developed specifically to mitigate these weaknesses~\cite{ref_RAG3}. RAG enhances an LLM by providing it with relevant, external information retrieved from a reliable knowledge base, as context for its generation task. Our preliminary investigations confirmed the necessity of this approach; directly prompting LLMs for path recommendation yielded inconsistent results. Therefore, we adopt the RAG architecture to ground and enhance the LLM's generation process in a factual database of historical paths, ensuring the recommended paths are both contextually relevant and plausible.\\

\begin{figure*} 
    \centering
    \includegraphics[scale=0.5]{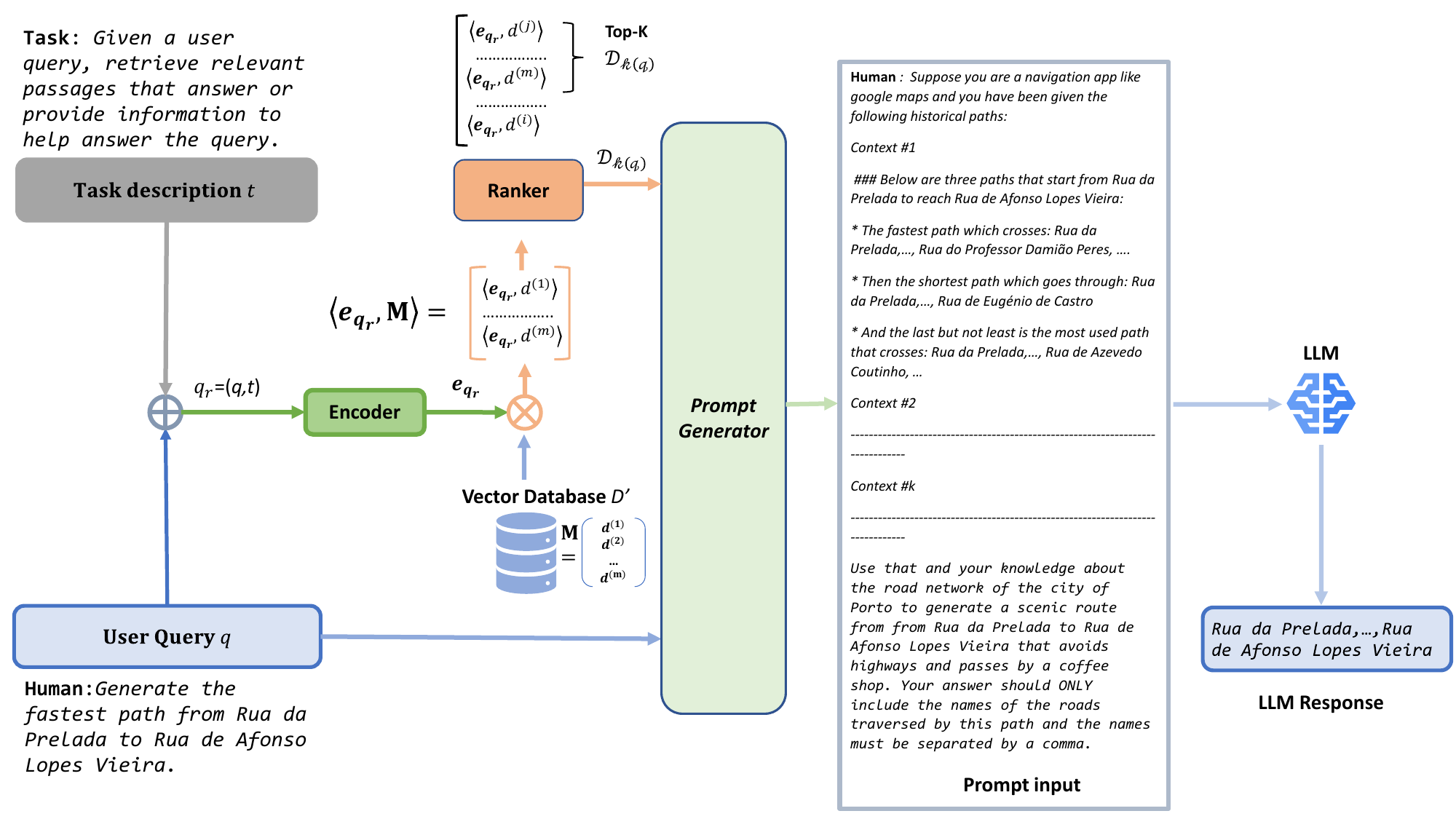} 
    \caption{PathGPT framework overview. PathGPT consists of two components, namely the context retrieval module and the generation module with the LLM at its core.}
    \label{fig:architecture} 
\end{figure*}

\section{Problem Formulation}

In this section, we formally define the problem of personalized path recommendation.

\begin{definition}
\textbf{Road network}. A road network is a directed graph ${G} = (\mathcal{V, E})$ where $\mathcal{V}$ is a set of nodes denoting the road intersections and $\mathcal{E} \subseteq \mathcal{V} \times \mathcal{V}$ is a set of edges representing road segments.   
\end{definition}

\begin{definition}
\textbf{Path}. A path is an ordered sequence of edges where 
$\mathcal{P} = \{e_1,...,e_{k - 1}\}$ with $e_i \in \mathcal{E}$ and $e_i = (v_i, v_{i+1}) \mbox{ for } i\in [1,...,k-1]$ and $v_i \in \mathcal{V}.$
\end{definition}

\begin{definition}
\textbf{Query}. The input query $q: (v_o, v_d, c)$ is a triple where $(v_o, v_d)$ denotes the OD pair (here we assume that $v_o, v_d \in \mathcal{V}$) and $c$ is a set of constraints that the path must satisfy.  
\end{definition}

\begin{definition}
\textbf{Path Recommendation Task}. Given a road network ${G}$, a query $q: (v_o, v_d, c)$, we would like to find a path $\mathcal{P^*}$ that starts from $v_o$ to reach $v_d$ while satisfying the given set of user constraints $c$ as much as possible.
\end{definition}

\section{Methodology}
\begin{figure*}
\centerline{
\includegraphics[scale=0.5]{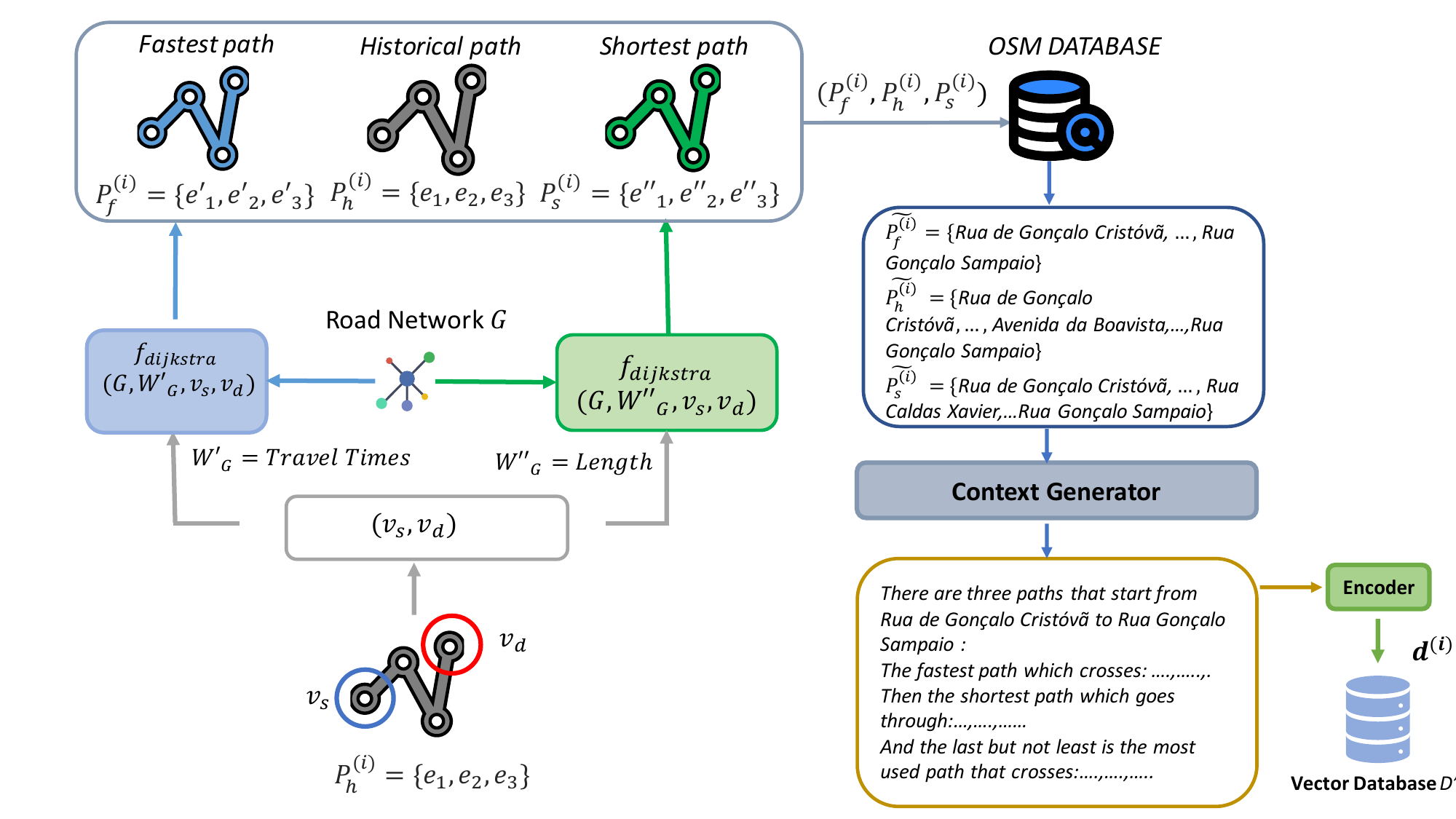}
}
\caption{The context generation process. We obtain the textual representations of a diverse set of paths.} 
\label{fig:context_generation}
\end{figure*}

This section details the architecture of PathGPT, a framework designed to apply LLMs to the path recommendation task. Our approach comprises a multi-stage pipeline that integrates data augmentation, a two-stage retrieval mechanism, and LLM-based generation. Initially, we construct a knowledge base by first augmenting the original trajectory data to create new path variations and then converting each path into a structured textual representation (Figure \ref{fig:context_generation}). During inference, when a user provides a query, this knowledge base is queried using our two-stage retrieval process to extract the \emph{top-k} most relevant paths. Finally, the textual descriptions of these paths are combined with the original query to create an augmented prompt, which instructs the LLM in generating the final, context-aware path (Figure \ref{fig:architecture}).

\subsection{Rationale for a Retrieval-Augmented Approach}
Our preliminary investigations into using off-the-shelf LLMs for path recommendation revealed a critical insight: while LLMs possess a nascent ability to reason about spatial navigation, their standalone performance is insufficient for reliable deployment. We identified one primary failure mode, which is generation inconsistency, where identical user prompts would sometimes result in significantly different paths. To address this, we adopted a strategy inspired by Retrieval-Augmented Generation (RAG). The central idea is to augment the user's query with relevant context retrieved from a database of historical trajectories. This retrieved information, which contains plausible paths that share similar origins, destinations, is designed to serve as a factual anchor, guiding the LLM's generation process and substantially improving the plausibility and consistency of its output.

\subsection{Knowledge Base Construction: From Trajectories to Text}
A core component of our methodology is the creation of a textual knowledge base from raw trajectory data. The historical paths in our initial dataset are represented as sequences of edge IDs, derived by map-matching each sequence of GPS coordinates data against the OpenStreetMap (OSM) road network~\cite{osm}. While these edge IDs are efficient machine identifiers, they are semantically opaque to an LLM; in this context, a sequence of numbers provides no useful information for generating a natural language-like path description.

To bridge this gap, we transform each sequence of edge IDs into a structured textual description. This transformation is possible because each OSM edge ID acts as a key to a rich database of road segment attributes. For each path, we perform lookups to retrieve salient information, such as official road names and lengths. We then synthesize this information into a coherent paragraph that includes:

\begin{itemize}
    \item The human-readable starting and destination addresses (obtained via reverse geocoding).
    \item The sequence of traversed road names.
    \item The specific type of the path (e.g., historical, fastest, shortest).

\end{itemize}

Furthermore, to enrich our knowledge base and improve the model's ability to generalize across different user objectives, we augment the dataset. For each historical path $\mathcal{P}$, we also compute and store two new, theoretically optimal paths: the fastest path and the shortest path $\mathcal{P}_f$ and $\mathcal{P}_s$ respectively, as illustrated in Figure \ref{fig:context_generation}. These two paths will serve as the basis for the generation of more user-tailored paths, since any of these paths, in a sense, is an extension of the fastest and shortest paths. These two additional paths are given by:

\begin{equation}
    \mathcal{P}_f = f_{dijkstra}(G, W', v_o, v_d)
\end{equation}

\begin{equation}
    \mathcal{P}_s = f_{dijkstra}(G, W'', v_o, v_d)
\end{equation}

where $f_{dijkstra}$ is Dijkstra's algorithm, ${G}$ is the road network graph, $W'$ and ${W''}$ are the weights of the edges of ${G}$ expressed in terms of edge travel times and length, respectively, and $(v_o, v_d)$ is the origin-destination (OD) pair.

While these three paths—$\mathcal{P}$, $\mathcal{P}_f$, $\mathcal{P}_s$—share common endpoints, they represent distinct routing strategies (historical, fastest, and shortest) and therefore typically consist of different edge sequences. Rather than indexing each path variant individually, our key insight is to consolidate them into a single, comprehensive document. We use a template-based context generator to create a unified paragraph (also called a document in this context), denoted as $d$, that describes the characteristics of all three path types in a human-readable format. This consolidated document now encapsulates a rich, multi-faceted context for a single origin-destination pair.

\subsection{Two-Stage Hybrid Retrieval}

Given the scale of the knowledge base, an efficient and effective retrieval strategy is paramount. A naive lexical search, while capable of matching exact keywords like street names, is brittle and often fails to capture the broader semantic intent of a user's query. Conversely, a purely semantic search might miss documents with precise but uncommon keyword matches. To leverage the strengths of both paradigms, we implement a two-stage hybrid retrieval system.

\subsubsection{Lexical Filtering with BM25.}The first stage acts as a fast, lexical filter designed to rapidly narrow the search space. For this, we employ BM25~\cite{robertson1995okapi}, a well-established probabilistic ranking function widely used in information retrieval. BM25 excels at identifying documents that are lexically relevant to a query by scoring them based on the term frequency (TF) and inverse document frequency (IDF) of the query terms. The process is as follows: both the user query $q$ and every document $d$ are tokenized. The tokens from the query are then used as keywords to score all documents in the corpus. This initial retrieval stage yields a candidate set of documents ranked by their BM25 relevance score, effectively filtering out the vast majority of irrelevant entries. For any document, its ranking score is given by:

\begin{equation}
score(q, d) = \sum_{i=1}^{n} \text{IDF}(q_i) \cdot \frac{f(q_i, d) \cdot (k_1 + 1)}{f(q_i, d) + k_1 \cdot \left(1 - b + b \cdot \frac{|d|}{\text{avgdl}}\right)}
\end{equation}

where $f(q_i, d)$ is the frequency of the token $q_i$ in document $d$, $|d|$ is the length of document $d$, avgdl is the average document length in the database, $k1$, $b$ are hyperparameters and $\text{IDF}(q_i)$ is the inverse document frequency given by:

\begin{equation}
\text{IDF}(q_i) = \log\left( \frac{N - n(q_i) + 0.5}{n(q_i) + 0.5} + 1 \right)
\end{equation}

\subsubsection{Semantic Re-ranking}
The candidate set produced by BM25 is lexically relevant but may not be perfectly aligned with the semantic intent of the user's query (e.g., "a quiet scenic drive"). Therefore, the second stage performs a semantic re-ranking of this filtered set to identify the documents that best match the query's underlying meaning. This is achieved by comparing vector embeddings, a technique central to modern RAG systems~\cite{rag}. A crucial step in this stage is the preparation of the user query $q$ for semantic encoding. As shown in Figure \ref{fig:context_generation}, the initial query is first augmented by concatenating it with a task-specific instruction, $t$. This augmentation is necessary to align the input with the fine-tuning paradigm of our sentence-transformer model, ensuring optimal retrieval performance. The resulting augmented query, $q_a$, is then encoded into a vector embedding, $\mathbf{e_{q_a}}$. The core of the re-ranking process is the computation of similarity scores. Let the set of document embeddings from the BM25 stage be represented by a matrix $\mathbf{M}$ of size $k' \times l$ where $k'$ is the number of candidates and $l$ is the embedding dimension. Thus, each row $i$ of this matrix is a pre-computed document vector $\mathbf{d^{(i)}}$. The similarity score $s(i)$ between the augmented query embedding $\mathbf{e_{q_a}}$ and each document embedding is then computed. 
\begin{equation}
s(i) = sim(\mathbf{e_{q_a}}, \mathbf{M_i}) \
\end{equation}

Where $sim$ is the cosine similarity metric given by:
\begin{equation}
sim(\mathbf{x}, \mathbf{y}) = \frac{\mathbf{x} \cdot \mathbf{y}}{\|\mathbf{x}\| \|\mathbf{y}\|}
\end{equation}

and $\mathbf{M_i}$ is the $i$-th column of $\mathbf{M}$.
After calculating the scores for all $k'$ candidates, they are ranked, and the top-$k$ documents are selected to serve as context for the LLM prompt.
\begin{equation} 
\mathcal{D}_{k(q_a)} =  \{d^{(j)},...,d^{(i)}\}
\end{equation}
where $\mathcal{D}_{k(q)}$ is the set of retrieved textual representations with respect to the augmented query $q_a$. In the next stage, we concatenate $\mathcal{D}_{k(q_a)}$ with $q_a$ to generate the final LLM prompt, as shown in Figure \ref{fig:architecture}.

\begin{equation} 
\mathcal{\hat{P}} =  LLM(q_a|\mathcal{D}_{k(q_a)})
\end{equation}

For reasons similar to the retriever model, we structure our prompt so that the first part describes the generation task to the LLM, then provides the context, and the task itself.

\section{Experiments}

\begin{table}[h]
\caption{Datasets details. We list the number of nodes and edges in the road network graph associated with each city. The initial number of historical paths from each dataset is also given.}\label{tab1}
\centering
\begin{tabularx}{\columnwidth}{XXXXXX}
\toprule
\textbf{ Dataset} & \textbf{ Nodes} & \textbf{Edges} & \textbf{Paths} & \textbf{RAG Corpus} \\ 
\midrule
Beijing &  31,199 & 72,156  & 1,382,948 & 4705\\  
Chengdu &  3,973 & 9,255 & 3,600,503 &  64657\\
Harbin  &  6,598 & 16,292 & 1,133,548 & 6646\\
\bottomrule
\end{tabularx}
\end{table}

\subsection{Experimental Setup}
\subsubsection{Datasets.}
We conduct offline experiments on publicly available historical taxi trajectory datasets ~\cite{dataset} from three different cities, namely Beijing (BJG), Chengdu (CHG), and Harbin (HRB). In Table \ref{tab1} we present the details of the datasets. Initially, each trajectory is a sequence of GPS coordinates; therefore, a map matching algorithm~\cite{map_matching} was applied to transform it to a sequence of edge IDs used to create the knowledge base. A small portion of the original dataset is also used to prompt the LLM. To maintain fairness during the evaluation, we exclude it from the knowledge base.

\subsubsection{Ground Truth Generation.}
We evaluated the performance of PathGPT on two path recommendation tasks: generating the most scenic path, the most fuel-efficient path for a given origin-destination (OD) pair. The ground truth datasets used during these evaluations are the following:

\begin{itemize}
    \item Scenic path: Generated by re-weighting the road network graph to reward proximity to points of interest (POIs). After identifying all POIs categorized as 'attractions,' 'leisure,' or 'amenities,' we applied a substantial weight discount to any road edge within a predefined proximity threshold of a POI. The resulting "shortest" path on this modified graph produces a path that balances directness with exposure to scenic locations.
    \item Fuel-Efficient Path: We devised a heuristic based on the principle that fuel efficiency is inversely related to high-speed travel due to factors like increased aerodynamic drag. Accordingly, we assigned a high penalty to any edge that was tagged as a 'highway' or had a maximum speed limit exceeding a predefined threshold. The "shortest" path computed on this weight-adjusted graph was then designated as the most fuel-efficient path.
    % \item Most Likely Path: Directly derived from our historical trajectory data, we define the most frequently traversed unique path in the dataset as the most likely path.\\
\end{itemize}

\subsubsection{Evaluation Metrics}
We evaluate the accuracy of PathGPT through two popular metrics: recall and precision, which are given by:
\begin{equation}
Precision= \cfrac{|\mathcal{P}^* \cap \mathcal{\hat{P}}|}{|\mathcal{P}^*|}
\end{equation}
\begin{equation}
    Recall = \cfrac{|\mathcal{P}^* \cap \mathcal{\hat{P}}|}{|\mathcal{\hat{P}}|}
\end{equation}
Here $\mathcal{P^*}$, $\mathcal{\hat{P}}$ represent the ground truth and the LLM-generated path, respectively. The generative nature of LLMs necessitates a robust validation system to prevent "hallucinations" and ensure that every recommended path is both valid and safe. A safety-critical application like navigation cannot tolerate the generation of disconnected, illegal, or non-existent paths. To guarantee reliability, we propose a post-processing validation layer where the LLM's natural language is parsed to extract the sequence of road names. Each name will then be mapped to its corresponding unique edge ID in the underlying road network graph (e.g., OpenStreetMap), resolving any ambiguities.
 
\subsubsection{Implementation Details.}
We use the shortest path algorithm implementation from Networkx~\cite{networkx} to compute both $\mathcal{P}_f$ and $\mathcal{P}_s$. We initially employ the multilingual-e5-large ~\cite{wang2024multilinguale5textembeddings} embedding model for encoding each document $d$ and the augmented query $q_a$. As LLM, we use the 4-bit quantized version of qwen2.5-14b-instruct~\cite{qwen}. Our experiments are conducted on a  machine running Ubuntu 22.04 LTS with Intel(R) Xeon(R) CPU @ 2.10GHz and one RTX 4090 GPU card.

\subsubsection{Baselines.}We compare the performance of our framework with the following models:
\begin{itemize}[label=$\bullet$]
    \item Shortest Path (SP): Computed using Dijkstra's algorithm with edges' weights set to edges' lengths.
    \item Fastest Path (FP):  Computed using Dijkstra's algorithm with edges' weights set to edges' travel times.
    \item NeuroMLR-D~\cite{neuromlr}: Trains a graph neural network on historical data using Lipschitz embeddings to generate paths. This is a version of NeuroMLR that uses Dijkstra's algorithm as a backend.
    \item NeuroMLR-G~\cite{neuromlr}: This is a version of NeuroMLR that uses a greedy algorithm.
    \item Base LLM w/o context: An off-the-shelf LLM without the context provided by PathGPT's retriever system.
\end{itemize}

\subsection{Experimental Results}

\subsubsection{Results Overview}The evaluation results for both recommendation tasks are presented in Tables \ref{tab4} and \ref{tab3}. A primary finding from our experiments is the profound impact of our retrieval-augmented framework. Across all evaluated tasks, PathGPT demonstrates a substantial performance improvement over the base LLM used without any retrieved context. This performance gap is often dramatic, with improvements exceeding 60\% in some cases. This provides strong empirical evidence for our core hypothesis: grounding the LLM with relevant, in-context examples is essential for generating accurate and plausible paths.

When compared with the specialized, learning-based baselines NeuroMLR-D and NeuroMLR-G, PathGPT exhibits a compelling performance profile that highlights its flexibility. Our framework generally outperforms these models on the scenic and fuel-efficient path generation tasks. Conversely, the NeuroMLR models, which were purpose-built and trained exclusively on historical data, achieve superior performance on some datasets. This outcome is expected and underscores the central advantage of PathGPT. While specialized models can excel at a single, predefined objective, PathGPT remains highly competitive across a diverse set of tasks, including those unseen during any training phase, without requiring any model retraining or fine-tuning. This demonstrates its significant potential as a generalizable and adaptive path recommendation system.

\subsubsection{Effect of the number of retrieved documents (\emph{k})}
To analyze the impact of context quantity on performance, we conducted a sensitivity analysis on the number of retrieved documents, $k$. We evaluated PathGPT using k values of 3, 6, and 9.
Our results show a clear positive correlation between the amount of retrieved context and performance within this tested range. Performance consistently improved as $k$ increased, with the optimal results in our analysis achieved at $k=9$. This suggests that providing a richer set of in-context examples allows the LLM to make more informed and accurate generations. However, we assume that this trend will not continue indefinitely. The performance is subject to a trade-off involving at least two key constraints. 

First, the finite context window of the LLM imposes a hard limit on the number of documents that can be processed, with larger values of $k$ increasing computational overhead. Second, and more critically, increasing $k$ raises the probability of retrieving less relevant or noisy documents. Such information could confuse the model and ultimately degrade, rather than enhance, performance. Therefore, identifying the optimal value for $k$ is a crucial tuning step that balances providing sufficient context against the risk of introducing noise.

\begin{table}[htbp]
\caption{Comparison of PathGPT against the baseline models on the precision and recall metrics on the three datasets for the scenic path recommendation task.}
\label{tab4}
\renewcommand{\arraystretch}{1.2}
\setlength{\tabcolsep}{4pt}
\begin{tabularx}{\columnwidth}{l|XXX|XXX}
\toprule
\multirow{2}{*}{\textbf{Models}} & \multicolumn{3}{c|}{\textbf{Precision (\%)}} & \multicolumn{3}{c}{\textbf{Recall (\%)}} \\
\cline{2-7}
& \textbf{BJG} & \textbf{CHG} & \textbf{HRB} & \textbf{BJG} & \textbf{CHG} & \textbf{HRB} \\
\midrule
SP      & 79.92   & 53.26   & 83.34          & \textbf{87.01}    & 61.72  & 87.66   \\
FP      & 78.47   & 51.23   & 81.44          & \underline{86.40}    & 61.52  & 86.33   \\
NeuroMLR-D      & 78.10   & 56.88   & 84.25          & 76.22    & 53.50  & 84.20   \\
NeuroMLR-G      & 77.49   & 56.67   & 78.87          & 76.15    & 53.73  & 82.42   \\
Base LLM & 49.94   & 30.14   & 46.63          & 32.84    & 29.14  & 25.65  \\
PathGPT@3       & 82.63   & 88.34   & 84.36          & 84.56   & \underline{92.46}  & 86.85  \\
PathGPT@6       & \underline{84.19}   & \underline{88.52}   & \underline{86.82}          & 84.13    & 91.61  & \underline{87.94}  \\
PathGPT@9       & \textbf{84.21}   & \textbf{89.85}   & \textbf{88.32}          & 84.60    & \textbf{92.91}  & \textbf{88.29}  \\
\bottomrule
\end{tabularx}
\end{table}

\begin{table}[htbp]
\caption{Comparison of PathGPT against the baseline models on the precision and recall metrics on the three datasets for the most fuel-efficient path recommendation task. The best results are written in bold, while the second-best are underlined.}
\label{tab3}
\renewcommand{\arraystretch}{1.2}
\setlength{\tabcolsep}{4pt}
\begin{tabularx}{\columnwidth}{l|XXX|XXX}
\toprule
\multirow{2}{*}{\textbf{Models}} & \multicolumn{3}{c|}{\textbf{Precision (\%)}} & \multicolumn{3}{c}{\textbf{Recall (\%)}} \\
\cline{2-7}
& \textbf{BJG} & \textbf{CHG} & \textbf{HRB} & \textbf{BJG} & \textbf{CHG} & \textbf{HRB} \\
\midrule
SP      & \underline{89.58}   & 78.61   & 83.28          & \textbf{90.54}    & 80.11  & 84.32   \\
FP      & 88.69   & 74.85   & 82.41          & \underline{90.13}    & 77.21  & 84.46   \\
NeuroMLR-D      & \textbf{89.08}   & 85.46   & 84.80          & 89.28    & 85.15  & 85.40 \\
NeuroMLR-G      & 87.31   & 85.14   & 80.66          & 88.16    & 84.91  & 83.21   \\
Base LLM & 54.72   & 35.29   & 50.59          & 31.63    & 26.06  & 26.89  \\
PathGPT@3       & 86.71   & 89.74   & 84.91          & 86.91    & \underline{92.78}  & 85.87  \\
PathGPT@6       & 87.81   & \underline{90.39}   & \underline{86.44}          & 87.48    & 92.61  & \textbf{86.56}  \\
PathGPT@9       & 88.39   & \textbf{90.79}   & \textbf{86.52}          & 88.38    & \textbf{92.81}  & \underline{86.17}  \\
\bottomrule
\end{tabularx}
\end{table}

\subsubsection{Processing Time} We report the average processing time of the retriever system and the inference time of the LLM in Figure \ref{fig:latency_datasets}. Across the three datasets, the average time taken to retrieve 9 documents from the corpus is around 3 seconds using bm25s' implementation~\cite{bm25s} of BM25 and Hugging Face~\cite{hf} to load the embedding model. The average LLM inference time is around 1.5 seconds when using Ollama~\cite{Ollama2025} as the inference backend. 

\begin{figure}[htbp]
\centering
\begin{tikzpicture}
\begin{axis}[
    ybar stacked,
    ymin=0,
    bar width=25pt,
    nodes near coords,
    enlargelimits=0.25,
    legend style={at={(0.5,-0.20)},
      anchor=north,legend columns=-1},
    ylabel={Processing Time (ms)},
    symbolic x coords={Beijing, Chengdu, Harbin},
    xtick=data,
    x tick label style={rotate=45,anchor=east}
    % width=0.45\textwidth,
    % height=0.36\textwidth
    ]
% Retrieval (bottom)
\addplot+[ybar] plot coordinates {
    (Beijing,2750) (Chengdu,3190) (Harbin, 3010)};

% Inference (stacked on top)
\addplot+[ybar] plot coordinates {
    (Beijing,1570) (Chengdu,1370) (Harbin,1660)};
  
\legend{\strut Retrieval Time, \strut LLM Inference Time with Ollama as Backend}
\end{axis}
\end{tikzpicture}
\caption{PathGPT's average processing time across three different real-world trajectory datasets. The bars show the time consumed by retrieval versus LLM inference.}
\label{fig:latency_datasets}
\end{figure}

\section{Path to Deployment and Future Work}
This work establishes that LLMs offer a  new paradigm for solving complex problems in routing and navigation. While our experiments demonstrate the significant potential of PathGPT as a flexible and adaptive path recommendation framework, its transition from a research prototype to a production-ready system requires addressing key challenges in computational efficiency, reliability, and user-centric validation. This section outlines a clear path toward making PathGPT a practical and trustworthy application.
\subsection{Addressing Latency and Scalability}
A primary consideration for any real-world navigation tool is near-instantaneous response time. Our current implementation has an average latency of approximately 4.5 seconds, which, while acceptable for offline experiments, is a bottleneck for on-the-fly path generation. Our strategy to address this is twofold:
\begin{itemize}
    \item Optimizing the Retrieval Pipeline: The retrieval stage currently accounts for the majority of the latency (~3 seconds). We plan to implement several standard industry optimizations. First, we will transition to a more efficient vector index, such as Hierarchical Navigable Small World (HNSW), to accelerate semantic search. Second, we can implement a caching layer for retrieval results, storing the context for frequently queried origin-destination pairs to bypass the search process entirely for common requests.
    \item Accelerating LLM Inference: To reduce the LLM's inference time (~1.5 seconds), we will explore model optimization techniques. While we already use a 4-bit quantized model, further improvements can be achieved through knowledge distillation. This involves training a smaller, specialized "student" model to replicate the path generation capabilities of the larger "teacher" model. A distilled model would have a significantly smaller memory footprint and lower computational cost, making it more suitable for deployment on edge devices or in resource-constrained server environments.
\end{itemize}
By addressing these challenges, we are confident that the novel paradigm presented by PathGPT can evolve into a powerful, reliable, and truly user-centric navigation service.

\subsection{Ensuring Reliability and Safety}

The generative nature of LLMs necessitates a robust validation system to prevent "hallucinations" and ensure that every recommended path is both valid and safe. A safety-critical application like navigation cannot tolerate the generation of disconnected, illegal, or non-existent paths. To guarantee reliability, we propose a post-processing validation layer:
Parsing and Entity Resolution: The LLM's natural language output will be parsed to extract the sequence of road names. Each name will then be mapped to its corresponding unique edge ID in the underlying road network graph (e.g., OpenStreetMap), resolving any ambiguities.
Path Connectivity and Rule Verification: Once the path is represented as a sequence of graph edges, we will perform two checks. First, we will verify that the sequence is contiguous—that each edge connects to the next. 

Second, we will validate the path against the graph's metadata, ensuring it adheres to all traffic laws, such as one-way street directions, turn restrictions, and vehicle access limitations.
Fallback Mechanism: If the generated path fails validation at any stage, the system will not return an error to the user. Instead, it will automatically fall back to a robust, conventional algorithm. For instance, if the query was for a "scenic path" and the LLM's output was invalid, the system could default to providing the standard fastest path (computed via Dijkstra's algorithm) along with a message informing the user. This ensures the user experience is never broken.

\subsection{Human-in-the-Loop for True Personalization}
Our current evaluation relies on heuristic-based ground truths for complex criteria like "scenic." A truly deployed system must learn from the subjective preferences of its users. To achieve this, we plan to incorporate a human-in-the-loop feedback mechanism. By allowing users to rate the generated paths (e.g., via a simple "thumbs up/down" interface), we can collect valuable preference data. This data can then be used to further fine-tune the LLM using techniques like Reinforcement Learning from Human Feedback (RLHF), progressively aligning the model's understanding of concepts like "scenic" or "quiet" with genuine human perception.
By systematically addressing these challenges, we are confident that the novel paradigm presented by PathGPT can evolve into a powerful, reliable, and truly user-centric navigation service.

\section{Conclusion}
This work establishes that LLMs offer a powerful and flexible new paradigm for solving complex problems in routing and navigation by proposing PathGPT, a framework leveraging a Retrieval-Augmented Generation (RAG) approach. Our experiments demonstrate that PathGPT successfully generates plausible paths that adhere to complex, user-specific constraints expressed in natural language, overcoming a key limitation of prior, inflexible models, while performing incredibly well compared to \emph{state of the art} models. While this approach is promising, its path to widespread adoption requires addressing critical challenges. Future research could focus on enhancing reliability and improving computational efficiency through model optimization techniques to further explore their potential on the PR task.

\bibliographystyle{unsrtnat}
\bibliography{main}  %%% Uncomment this line and comment out the ``thebibliography'' section below to use the external .bib file (using bibtex) .

\begin{thebibliography}{30}
\providecommand{\natexlab}[1]{#1}
\providecommand{\url}[1]{\texttt{#1}}
\expandafter\ifx\csname urlstyle\endcsname\relax
  \providecommand{\doi}[1]{doi: #1}\else
  \providecommand{\doi}{doi: \begingroup \urlstyle{rm}\Url}\fi

\bibitem[Jain et~al.(2021)Jain, Bagadia, Manchanda, and Ranu]{neuromlr}
J.~Jain, V.~Bagadia, S.~Manchanda, and S.~Ranu.
\newblock Neuromlr: Robust and reliable route recommendation on road networks.
\newblock In \emph{Proc. 35th Conf. Neural Inf. Process. Syst.}, volume~34,
  pages 22070--22082, 2021.

\bibitem[Jiang et~al.(2022)Jiang, Dong, Wu, and Liu]{increase_net}
Zhuojun Jiang, Lei Dong, Lun Wu, and Yu~Liu.
\newblock Quantifying navigation complexity in transportation networks.
\newblock \emph{PNAS Nexus}, 1\penalty0 (3):\penalty0 pgac126, 7 2022.

\bibitem[Quercia et~al.(2014)Quercia, Schifanella, and Aiello]{personal_route}
D.~Quercia, R.~Schifanella, and L.~M. Aiello.
\newblock The shortest path to happiness: Recommending beautiful, quiet, and
  happy routes in the city.
\newblock In \emph{Proc. 25th ACM Conf. Hypertext Soc. Media}, pages 116--125,
  Santiago, Chile, 2014. ACM.

\bibitem[Ceikute and Jensen(2013{\natexlab{a}})]{driver_behavior}
V.~Ceikute and C.~S. Jensen.
\newblock Routing service quality: Local driver behavior versus routing
  services.
\newblock In \emph{Proc. 14th IEEE Int. Conf. Mobile Data Manage.}, pages
  97--106. IEEE, 2013{\natexlab{a}}.

\bibitem[Ceikute and Jensen(2013{\natexlab{b}})]{leisure_drive}
V.~Ceikute and C.~S. Jensen.
\newblock Routing service quality: Local driver behavior versus routing
  services.
\newblock In \emph{Proc. 14th IEEE Int. Conf. Mobile Data Manage.}, volume~1,
  pages 97--106, Milan, Italy, 2013{\natexlab{b}}. IEEE Comput. Soc.

\bibitem[Wang et~al.(2017)Wang, Chen, Wu, and Xiong]{gps_data_1}
J.~Wang, C.~Chen, J.~Wu, and Z.~Xiong.
\newblock No longer sleeping with a bomb: A duet system for protecting urban
  safety from dangerous goods.
\newblock In \emph{Proc. 23rd ACM SIGKDD Int. Conf. Knowl. Discov. Data
  Mining}, pages 1673--1681. ACM, 2017.

\bibitem[Li et~al.(2020)Li, Cong, and Cheng]{prr_1}
X.~Li, G.~Cong, and Y.~Cheng.
\newblock Spatial transition learning on road networks with deep probabilistic
  models.
\newblock In \emph{Proc. 36th IEEE Int. Conf. Data Eng.}, pages 349--360. IEEE,
  2020.

\bibitem[Brown et~al.(2020)Brown, Mann, Ryder, Subbiah, Kaplan, Dhariwal,
  Neelakantan, Shyam, Sastry, Askell,
  et~al.]{brown2020languagemodelsfewshotlearners}
T.~B. Brown, B.~Mann, N.~Ryder, M.~Subbiah, J.~Kaplan, P.~Dhariwal,
  A.~Neelakantan, P.~Shyam, G.~Sastry, A.~Askell, et~al.
\newblock Language models are few-shot learners.
\newblock \emph{arXiv preprint}, 2020.

\bibitem[Delling et~al.(2013)Delling, Goldberg, Nowatzyk, and
  Werneck]{ref_Dijkstra}
D.~Delling, A.~V. Goldberg, A.~Nowatzyk, and R.~F. Werneck.
\newblock Phast: Hardware-accelerated shortest-path trees.
\newblock \emph{J. Parallel Distrib. Comput.}, 73\penalty0 (7):\penalty0
  940--952, 2013.

\bibitem[Hart et~al.(1968)Hart, Nilsson, and Raphael]{ref_A}
P.~E. Hart, N.~J. Nilsson, and B.~Raphael.
\newblock A formal basis for the heuristic determination of minimum cost paths.
\newblock \emph{IEEE Trans. Syst. Sci. Cybern.}, 4\penalty0 (2):\penalty0
  100--107, 1968.

\bibitem[Liu and Zhang(2022)]{ref_IAFC}
T.~Liu and J.~Zhang.
\newblock An improved path planning algorithm based on fuel consumption.
\newblock \emph{J. Supercomput.}, pages 1--31, 2022.

\bibitem[Zhong et~al.(2020)Zhong, Tian, Hu, and Peng]{ref_A-DWA}
X.~Zhong, J.~Tian, H.~Hu, and X.~Peng.
\newblock Hybrid path planning based on safe a* algorithm and adaptive window
  approach for mobile robot in large-scale dynamic environment.
\newblock \emph{J. Intell. Robot. Syst.}, 99\penalty0 (1):\penalty0 65--77,
  2020.

\bibitem[Huang and Chang(2022)]{ref_LLM}
J.~Huang and K.~C. Chang.
\newblock Towards reasoning in large language models: A survey.
\newblock \emph{arXiv preprint}, 2022.

\bibitem[{DeepSeek-AI} et~al.(2025){DeepSeek-AI}, Guo, Yang, Zhang,
  et~al.]{ref_deepseek}
{DeepSeek-AI}, D.~Guo, D.~Yang, H.~Zhang, et~al.
\newblock Deepseek-r1: Incentivizing reasoning capability in llms via
  reinforcement learning.
\newblock \emph{arXiv preprint}, 2025.

\bibitem[Brown(2020)]{ref_LLM_benefit1}
T.~B. Brown.
\newblock Language models are few-shot learners.
\newblock \emph{arXiv preprint}, 2020.

\bibitem[Wei et~al.(2022)Wei, Wang, Schuurmans, Bosma, Xia, Chi, Le, and
  Zhou]{ref_LLM_benefit2}
J.~Wei, X.~Wang, D.~Schuurmans, M.~Bosma, F.~Xia, E.~Chi, Q.~V. Le, and
  D.~Zhou.
\newblock Chain-of-thought prompting elicits reasoning in large language
  models.
\newblock In \emph{Proc. 36th Conf. Neural Inf. Process. Syst.}, volume~35,
  pages 24824--24837, 2022.

\bibitem[He et~al.(2023)He, Zhang, and Roth]{ref_RAG1}
H.~He, H.~Zhang, and D.~Roth.
\newblock Rethinking with retrieval: Faithful large language model inference.
\newblock \emph{arXiv preprint}, 2023.
\newblock Version 2.

\bibitem[Ji et~al.(2023)Ji, Lee, Frieske, Yu, Su, Xu, Ishii, Bang, Madotto, and
  Fung]{ref_RAG2}
Z.~Ji, N.~Lee, R.~Frieske, T.~Yu, D.~Su, Y.~Xu, E.~Ishii, Y.~J. Bang,
  A.~Madotto, and P.~Fung.
\newblock Survey of hallucination in natural language generation.
\newblock \emph{ACM Comput. Surv.}, 55\penalty0 (12):\penalty0 1--38, 3 2023.

\bibitem[Ren et~al.(2023)Ren, Wang, Qu, Zhao, Liu, Tian, Wu, Wen, and
  Wang]{ref_RAG3}
R.~Ren, Y.~Wang, Y.~Qu, W.~X. Zhao, J.~Liu, H.~Tian, H.~Wu, J.~R. Wen, and
  H.~Wang.
\newblock Investigating the factual knowledge boundary of large language models
  with retrieval augmentation.
\newblock \emph{arXiv preprint}, 2023.

\bibitem[{OpenStreetMap contributors}(2024)]{osm}
{OpenStreetMap contributors}.
\newblock Openstreetmap researcher information.
\newblock Online, 2024.
\newblock URL \url{https://wiki.openstreetmap.org/wiki/Researcher_Information}.
\newblock Accessed: 2024-07-30.

\bibitem[Robertson et~al.(1995)Robertson, Walker, Jones, Hancock-Beaulieu, and
  Gatford]{robertson1995okapi}
Stephen~E. Robertson, Steve Walker, Susan Jones, Micheline Hancock-Beaulieu,
  and Mike Gatford.
\newblock Okapi at trec-3.
\newblock In Donna~K. Harman, editor, \emph{Proceedings of the Third Text
  REtrieval Conference (TREC-3)}, volume 500-225 of \emph{NIST Special
  Publication}, pages 109--126, Gaithersburg, MD, 1995. National Institute of
  Standards and Technology.

\bibitem[Lewis et~al.(2020)Lewis, Perez, Piktus, Petroni, Karpukhin, Goyal,
  Küttler, Lewis, tau Yih, Rocktäschel, Riedel, and Kiela]{rag}
P.~Lewis, E.~Perez, A.~Piktus, F.~Petroni, V.~Karpukhin, N.~Goyal, H.~Küttler,
  M.~Lewis, Wen tau Yih, T.~Rocktäschel, S.~Riedel, and D.~Kiela.
\newblock Retrieval-augmented generation for knowledge-intensive nlp tasks.
\newblock \emph{Proc. 34th Conf. Neural Inf. Process. Syst.}, pages 9459--9474,
  2020.
\newblock NeurIPS 2020 Oral Presentation.

\bibitem[Lian and Zhang(2018)]{dataset}
J.~Lian and L.~Zhang.
\newblock Beijing taxi gps trajectory dataset with vehicle status annotations.
\newblock In \emph{Proc. 1st Workshop Data Acquis. Anal.}, pages 3--4. ACM,
  2018.

\bibitem[Yang and Gidofalvi(2018)]{map_matching}
C.~Yang and G.~Gidofalvi.
\newblock Fast map matching: An algorithm integrating precomputation with
  hidden markov models.
\newblock \emph{Int. J. Geogr. Inf. Sci.}, 32\penalty0 (3):\penalty0 547--570,
  2018.

\bibitem[Hagberg et~al.(2008)Hagberg, Schult, and Swart]{networkx}
A.~A. Hagberg, D.~A. Schult, and P.~J. Swart.
\newblock Exploring network structure, dynamics, and function using {NetworkX}.
\newblock In G.~Varoquaux, T.~Vaught, and J.~Millman, editors, \emph{Proc. 7th
  Python Sci. Conf.}, pages 11--15, Pasadena, CA, 8 2008. SciPy.

\bibitem[Wang et~al.(2024)Wang, Yang, Huang, Yang, Majumder, and
  Wei]{wang2024multilinguale5textembeddings}
Liang Wang, Nan Yang, Xiaolong Huang, Linjun Yang, Rangan Majumder, and Furu
  Wei.
\newblock Multilingual e5 text embeddings: A technical report, 2024.
\newblock URL \url{https://arxiv.org/abs/2402.05672}.

\bibitem[Yang et~al.(2024)Yang, Yang, Hui, et~al.]{qwen}
A.~Yang, B.~Yang, B.~Hui, et~al.
\newblock {Qwen2} technical report.
\newblock \emph{arXiv preprint}, 2024.
\newblock Version 1.0.

\bibitem[Lù(2024)]{bm25s}
Xing~Han Lù.
\newblock Bm25s: Orders of magnitude faster lexical search via eager sparse
  scoring, 2024.
\newblock URL \url{https://arxiv.org/abs/2407.03618}.

\bibitem[HF(2025)]{hf}
HF.
\newblock Huggingface.
\newblock \url{https://huggingface.co/}, 2025.

\bibitem[Ollama(2025)]{Ollama2025}
Ollama.
\newblock Ollama.
\newblock \url{https://ollama.com}, 2025.

\end{thebibliography}




% 	\bibitem{keshet2016prediction}
% 	Keshet, Renato, Alina Maor, and George Kour.
% 	\newblock Prediction-Based, Prioritized Market-Share Insight Extraction.
% 	\newblock In {\em Advanced Data Mining and Applications (ADMA), 2016 12th International 
%                       Conference of}, pages 81--94,2016.

% \end{thebibliography}

\end{document}